# 16. Composition of solar system small bodies

Pierre Vernazza
*Aix Marseille Université and Laboratoire d'Astrophysique de Marseille*
Pierre Beck
*Université Grenoble Alpes and Institut de Planétologie et d'Astrophysique de Grenoble*

## 16.1 Introduction

Small bodies in our stellar system are rocky and/or icy objects, usually ranging in size from a few meters to a few hundreds of kilometers. They comprise main belt asteroids, giant planet Trojans, trans-Neptunian objects (TNOs), and comets. Their physical nature, distribution, formation, and evolution are fundamental to understanding how the solar system formed, evolved, and ultimately, why water and life are present on Earth.

'In the present day solar system, they are the most direct remnants of the original building blocks that formed the terrestrial planets and the solid cores of the giant planets. As such, they contain a relatively pristine record of the initial conditions that existed in our solar nebula some 4.6 Gyrs ago. The small bodies that have survived since that epoch, however, have experienced numerous collisional, dynamical, and thermal events that have shaped their present-day physical and orbital properties. Interpreting this record via observations, laboratory studies, and theoretical/numerical modeling can tell us much about the primordial state of these bodies and how they have evolved thereafter. In fact, even though small bodies represent only a tiny fraction of the total mass of the planets, their large numbers, diverse compositions, and orbital distributions provide powerful constraints for planetary system formation models' (Bottke et al. 2002).

Observations of the solar system small bodies performed mainly between the early 1970s and today led to the determination of orbits as well as the characterization of the visible and near-infrared spectral propeties, sizes and albedos for i) a statistically significant sample of small bodies up to 5 AU and ii) a restricted number of small bodies beyond 5 AU. Inter alia, these datasets have allowed us to reach a preliminary understanding of the compositional distribution across the different populations of small bodies and of the architecture of our solar system. In particular, they have revealed a number of puzzling features in each population of small bodies (e.g., compositional overlap in the asteroid belt, the Jupiter Trojans' large range of orbital inclinations, the complex orbital architecture of the Kuiper belt, etc..).

These observations – coupled with those of giant extrasolar planets close to stars – were instrumental in imposing giant planet migrations as a main step in the dynamical evolution of the solar system. On the basis of these datasets, the idea of a static solar system history has dramatically shifted to one of dynamic change and mixing (DeMeo & Carry 2014).

Dynamical models (Grand Tack and Nice models ; Walsh et al. 2011; Gomes et al., 2005; Tsiganis et al., 2005; Morbidelli et al., 2005 ; Bottke et al. 2006, 2012 ; Levison et al. 2009) currently propose that today's asteroid belt may not only hosts objects that formed in situ, typically between 2.2 and 3.3 AU, but also bodies that were formed in the terrestrial planet region as well as beyond Jupiter (and for some of them even beyond Neptune). They also propose that a large fraction of both the Jupiter Trojans and the Hildas has formed in more distant regions - typically in the primordial trans-neptunian disk, which is also the precursor of the Kuiper belt.

The aim of the present chapter is first to summarize our present understanding of the compositional distribution across the different reservoirs of small bodies (main belt asteroids, giant planet trojans, irregular satellites of the giant planets, TNOs, comets). We then use this information to i) discuss the dynamical models cited above, ii) mention possible caveats in these models if any, and iii) draw a preliminary version of the primordial compositional gradient across the solar system before planetary migrations occured. Note that the composition of both planetary satellites (the regular ones) and that of the transient populations (NEOs, centaurs) is not discussed here. We strictly focus on the composition of the main reservoirs of small bodies. The manuscript's objective is to provide a global and synthetic view of small bodies' compositions rather than a very detailed one; for specific reviews regarding the composition of small bodies, see papers by Burbine (2014) for asteroids, Emery et al. (2015) for Jupiter trojans, Mumma and Charnley (2011) for comets, and Brown (2012) for KBOs.

**16.2 Main belt asteroids**

16.2.1 S-complex asteroids

16.2.1.1 Surface composition

S-complex asteroids (see DeMeo et al. 2009 for the most recent asteroid taxonomy) are the dominating population in the inner region (2AU<semi-major axis<2.5AU) of the asteroid belt (Gradie & Tedesco 1982; DeMeo & Carry 2013). Overall, they are the third most important main belt population, representing ~8% of its mass (DeMeo & Carry 2013). Early comparisons between meteorite and asteroid optical reflectance spectra (e.g., Chapman and Salisbury 1973) suggested that only S-type asteroids have spectral characteristics (e.g., absorption band near 0.95 micron) compatible with those seen in the most common type of meteorite to fall to Earth, namely ordinary chondrites (OCs; ~80% of the falls) although S-type asteroids have a redder spectral slope. Subsequent measurements in the near-infrared range have not altered this picture (e.g., Gaffey et al. 1993, Vernazza et al. 2008, Vernazza et al. 2009a, DeLeon et al. 2010, Dunn et al. 2013, Vernazza et al. 2014). The spectral slope difference between OC and S-type spectra has, however, been an obstacle for more than 30 years to establishing a definite link between both groups and has constituted one of the greatest conundrums in the asteroid-meteorite field (see Chapman (2004) for an extensive review on this subject). The conundrum was even reinforced on the basis of mineralogical arguments (Gaffey et al. 1993) that were suggesting that a large fraction of S-type asteroids are partly differentiated and thus not OC-like (Abell et al. 2007). This paradox has gradually been resolved over the years. It is now understood that space weathering processes are responsible for the spectral mismatch between S-type bodies and OCs (e.g., Sasaki et al. 2001, Strazzulla et al. 2005, Marchi et al. 2005, Brunetto et al. 2006, Vernazza et al. 2009a). Furthermore, both telescopic observations (Vernazza et al. 2008, 2014) and the first asteroid sample return mission (Hayabusa; Nakamura et al. 2011, Yurimoto et al. 2011, Nakashima et al. 2014, Tsuchiyama et al. 2014) indicate that most S-type bodies have mineralogies similar to those of OCs. Importantly, the S-type/OC link has been delivering fundamental constraints on the formation and evolution of planetesimals over the recent years (Vernazza et al. 2014, 2015a):

1) most S-type asteroids, including large ones (D≈100-200 km), though not members of one family, are distributed into two well-defined compositional groups, Hebe-like and Flora-like

(H-like and LL-like; see Fig. 16.1). This indicates that identical compositions among multiple asteroids are a natural outcome of planetesimal formation, making it possible that meteorites within a given class originate from multiple parent bodies;

2) the surfaces of nearly all of these asteroids (up to 200 km) show the same compositional characteristics as high temperature meteorites (type 4 to 6 OCs) that were thermally metamorphosed in their interiors, these exposed interiors being a likely consequence of impacts by small asteroids (D<10km) in the early solar system (Ciesla et al. 2013);

3) the lack of compositional variation within both H-like asteroid families and the surfaces of the family members showing the same compositional characteristics as high temperature meteorites (type 4 to 6 OCs) is consistent with their parent bodies having been metamorphosed throughout, which implies – following current thermal models - that their formation process must have been rapid. Note that such a short duration of accretion as implied by the observations of Vernazza et al. (2014) is consistent with current models of planet formation through streaming and gravitational instabilities (Youdin & Goodman 2005, Johansen et al. 2007, Chiang & Youdin 2010) that show that bodies of several hundred kilometers in size form on the time-scale of a few orbits (Johansen et al. 2011, Youdin 2011, Johansen et al. 2012);

4) LL-like bodies possibly formed slightly closer to the Sun than H-like bodies (Vernazza et al. 2014), a possible consequence of radial mixing and size sorting of chondrules in the protoplanetary disk prior to accretion;

Importantly, S-type asteroids are a diverse group of objects and one cannot exclude that some meteorites other than the ordinary chondrites could also originate from them (e.g., *Burbine et al., 2003*). Possibilities include differentiated meteorites, as proposed by *Sunshine et al.* (2004), such as pallasites, brachinites, ureilites, lodranites, acapulcoites, winonaites, IAB irons, and mesosiderites (*Gaffey et al., 1993*). In a recent review, Vernazza et al. (2015a) have shown that both lodranites and acapulcoites have spectral properties that are very similar to those of OCs and H chondrites in particular. As such, some S-type asteroids could indeed be linked to these two meteorite classes. The remaining achondritic classes, however, appear as poor analogues of S-type asteroid surfaces.

16.2.1.2 Bulk composition

Density estimates more directly inform us of the bulk composition. However, in the case of small bodies - asteroids in particular - densities are highly dependent on the collisional history. Specifically, most D<100km asteroids are seen as collisionaly evolved objects (Morbidelli et al. 2009) whose internal structure is largely filled by voids (up to 50-60% in some cases; Consolmagno et al. 2008; Carry 2012) thus limiting our capability to interpret meaningfully their bulk density in terms of composition(s). Larger bodies (D>100km), on the other hand, are seen as "primordial" (Morbidelli et al. 2009), and for these objects the macroporosity is supposed to be minimal. As such, their density gives us a much better estimate of their initial bulk composition. Accordingly, we will only mention densities for objects with D>100km throughout the manuscript to reflect as much as possible their bulk composition.

Binary systems provide, via the orbit of the satellite, the information necessary for a density to be calculated for the primary. So far, no binary system has been detected among large (D>100km) S-type asteroids. As such, density measurements for S-type asteroids are less precise than those acquired for instance for C-complex and P/D-type asteroids. As of today, there is no precise density estimate for the largest S-types (D>100km ; Carry 2012).

## 16.2.2 C-complex asteroids

### 16.2.2.1 Surface composition

C-complex asteroids, which comprise six subclasses (B, C, Cb, Cg, Cgh, Ch; DeMeo et al. 2009), are the dominating population in the asteroid belt, representing ~60% of its mass (DeMeo & Carry 2013, 2014). If one removes the four most massive asteroids (Ceres, Pallas, Vesta and Hygeia), C-complex asteroids still represent ~40% of the main belt mass (DeMeo & Carry 2013). They are mainly located in the middle (2.5-2.8 AU) and outer (2.8-3.3 AU) part of the belt. While one third of all C-complex asteroids (Ch- and Cgh-types; Rivkin 2012) possess a well-established meteoritic analog (CM chondrites; e.g., Burbine 1998; see Fig. 16.1), the bodies in the remaining subclasses (B-, C-, Cb- and Cg-type), including Ceres, appear unsampled by our meteorite collections (Vernazza et al. 2015b). Whereas metamorphosed CI/CM chondrites have been proposed in the past as analogues of B-, C-, Cb- and Cg-type surfaces (Hiroi et al. 1993), such possibility is presently untenable for the majority of these asteroids for three reasons reasons, namely the paucity of these meteorites among falls (~0.2% of meteorite falls) compared to the abundance of B-, C-, Cb- and Cg-types, the difference in density between these meteorites and these asteroids, and the difference in spectral properties (in the mid-infrared) between the two groups (see Vernazza et al. 2015b for discussion).

Vernazza et al. (2015b) have recently shown that B-, C-, Cb- and Cg-type asteroids (hereafter BCG-type asteroids) have spectral properties compatible with those of pyroxene-rich IDPs, opening the possibility that IDPs may sample a larger fraction of main belt asteroids (in terms of mass) than meteorites.

Finally, whereas Ch- and Cgh-types appear compositionally distinct from BCG-type asteroids, it is important to notice that both populations are concentrated in the same region of the asteroid belt (Bus 1999, Rivkin 2012).

### 16.2.2.2 Bulk composition

The high-precision density measurements collected so far for large binary C-complex asteroids indicate that Ch and Cgh-types (CM-like bodies) have significantly higher densities than BCG-types (IDP-like bodies). Specifically, Ch- and Cgh-types (41) Daphne, (121) Hermione and (130) Elektra have densities in the 1.9-2.4 g/cm$^3$ range whereas BCG-types (45) Eugenia, (90) Antiope, (253) Mathilde and (762) Pulcova have densities in the 0.8-1.5 g/cm$^3$ range (Marchis et al. 2012, Carry et al. 2012).

The few existing density measurements for BCG-types (the same applies for P- and D-types) thus indicate that these bodies cannot be made of silicates only, and must comprise a significant fraction of ices. This comet-like nature applies also to the dwarf planet Ceres as plumes of water vapour have unambiguously been detected at this object (Kuppers et al. 2014; Ceres'mass ~35% of the main belt mass). Ceres, with a density of ~2.13 g/cm$^3$ (Carry 2012) is expected to have a rock/ice ratio of about 1. It is thus not a surprise that several so-called main-belt comets were discovered among BCG-type objects (Hsieh & Jewitt 2006, Jewitt 2012). Additional support to the comet-like nature of these objects is provided by spectroscopy in the 3-micron region which indicates that water ice may not only be present in the interior of these bodies but also on their surfaces (Campins et al. 2010, Rivkin & Emery 2010, Takir & Emery 2012).

16.2.3 P- and D-type asteroids

16.2.3.1 Surface composition

P- and D-type asteroids represent ~11% of the main belt mass (DeMeo & Carry 2013; note : we call P-types both the low albedo (pv<0.1) X-types and T-types from the Bus taxonomy ; Bus 1999). Since both asteroid types are the two main populations of small bodies among Jupiter Trojans, we will not address their surface composition in this subsection but rather in the Jupiter Trojans' one (16.3).

16.2.3.2 Bulk composition

Up to now, two large (D>200km) P-type main belt asteroids have been found to have satellites and the derived densities are low (1.4 ± 0.2 and 1.5 ± 0.2 for 87 Sylvia and 107 Camilla respectively; Marchis et al. 2012). These low densities likely are consistent with the presence of ice(s) in the interior of these bodies.

16.2.4 Remaining asteroid types (A, K, L, M, O, R, V, Xe, Xc, Xk)

16.2.4.1 Surface composition

Excluding (4) Vesta, the asteroid types A, K, L, M, O, R, V, Xe, Xc and Xk represent less than ~7% of the main belt mass, which illustrates the rarity of these types in the asteroid belt (note : we call M-types the high albedo (pv>0.1) X-types from the Bus taxonomy ; Bus 1999). These types are the likely parent bodies of all meteorite classes but OCs and CMs, that is of ~18% of the meteorites among falls. Most of these meteorite classes are achondritic and as such most of these asteroid types are fragments of differentiated bodies when not quasi intact differentiated protoplanets such as Vesta. Specifically, ground-based observations currently suggest the following associations between asteroid and meteorite types (see Fig. 16.1):

- A types comprise the parent bodies of pallasites and brachinites (e.g., Cruikshank and Hartmann, 1984; Sunshine et al. 2007)
- K types comprise the parent bodies of CV, CO, CR, CK meteorites (e.g., Bell 1988, Burbine et al. 2001, 2002; Clark et al. 2009). Note that CVs might originate from a differentiated parent body (Weiss and Elkins-Tanton, 2013).
- M types comprise the parent bodies of iron meteorites (e.g., Cloutis et al., 1990; Shepard et al. 2015 and references therein)
- V types comprise the parent bodies of HED meteorites (e.g., Consolmagno and Drake, 1977)
- Xc types comprise the parent bodies of enstatite chondrite and aubrites (e.g., Zellner 1975, Zellner et al. 1977, Vernazza et al. 2009b, 2011, Shepard et al. 2015)
- Xk types comprise the parent bodies of mesosiderites (e.g., Vernazza et al. 2009b)

L-, R-, O- and Xe-types seem to be unsampled by our collections. Note that the R- and O-classes possess only one member so far (R: 349; O: 3628; Binzel et al. 1993, Bus 1999, DeMeo et al. 2009).

A number of L-types have absorption features consistent (Sunshine et al., 2008) with iron oxide-bearing aluminous spinel found in calcium–aluminum inclusions (CAIs; Burbine et al. 1992, Sunshine et al. 2008). Sunshine et al. (2008) argue that these objects contain extremely high (30±10 vol.%) CAI abundances compared to meteorites and may be more ancient than anything found in our meteorite collections. They suggest that these objects may have formed before the injection of the $^{26}$Al into the solar system since these objects would have melted with such high CAI abundances and canonical initial $^{26}$Al/$^{27}$Al ratios.

The R-type 349 Dembowska possesses a visible and near-infrared spectrum very similar to those of S-types with the exception that its absorption bands around 1 and 2 microns are significantly deeper than in the S-type cases (Bus & Binzel 2002, DeMeo et al. 2009). Both low-ca pyroxene and olivine dominate the surface mineralogy of 349 Dembowska (Hiroi & Sasaki 2001).

Until now, there has been no convincing interpretation of the surface composition of the O-type 3628 (Cloutis et al. 2006, Burbine et al. 2006, 2011).

As opposed to Xc-types, Xe-types do not have entirely featureless visible and near-infrared spectra. A number of weak absorption bands have been found in their reflectance spectra, including ~0.5, ~0.9 and ~1.8 μm bands (Bus and Binzel 2002, Kelley and Gaffey 2002, Clark et al. 2004a,b). The ~0.5 μm absorption has been argued to be due to oldhamite (CaS) by Burbine et al. (2002a), a mineral commonly found in aubrites and enstatite chondrites but usually in very low abundances. Weak bands at ~0.9 and ~1.8 μm have been attributed to a low-FeO pyroxene (Clark et al., 2004a; Gaffey et al., 1989).

16.2.4.2 Bulk composition

We have precise density measurements for two large asteroids belonging to these types as they have been visited by a spacecraft. (4) Vesta and (21) Lutetia have densities of respectively 3.46 and 3.4 g/cm$^3$ (Russell et al. 2012, Sierks et al. 2011).

**16.3 Jupiter Trojans**

16.3.1 Surface composition

Reflectance spectroscopy of Jupiter Trojans at visible and near-infrared wavelengths has failed to discover any absorption features, including no evidence for H$_2$O, no 1 and 2 μm silicate bands, and no absorptions from organics or hydrated minerals (Emery & Brown 2003, Emery et al. 2011). Instead, it has revealed red spectral slopes, comparable to main belt P- and D-type asteroids and cometary nuclei (see Fig. 16.2). No ultra-red slopes comparable to many Centaurs and TNOs have been detected among the Trojans, suggesting a different origin or different dynamical and physico-chemical evolutions. Emery et al. (2011) found evidence for two distinct spectral groups (P- and D-types) and attributed the difference in slope to two distinct compositions rather than to a diversity of regolith ages. Note that a few C-types are also present among Trojans (DeMeo & Carry 2013). In particular, some of these bodies have been found within the Eurybates family opening the possiblity that the interior of P- and D-type asteroids may consist of C-type material (De Luise et al. 2010).

At longer wavelengths, discrete mineralogical features attributed to fine-grained (~few μm), anhydrous silicates were detected in mid-IR thermal emission spectra of three Trojans using the Spitzer Space Telescope (Emery et al. 2006). The mineralogy resemble that of cometary silicates (see Fig. 16.2) and anhydrous IDPs (Emery et al. 2006, Vernazza et al. 2012, 2015b), and the spectral shape indicates that the surfaces are either very porous or that the grains are imbedded in a matrix that is transparent in the mid-IR (Emery et al. 2006, Vernazza et al. 2012, Yang et al. 2013). Recently, Vernazza et al. (2015b) have shown that the red Trojans (D-types) have spectral properties compatible with those of olivine-rich IDPs.

16.3.2 Bulk composition

Up to now, two Trojan asteroids have been found to have satellites and the derived densities are very low (1.1 ± 0.3 g/cm$^3$ and 1.0 ± 0.3 g/cm$^3$ for 617 Patroclus and 624 Hektor respectively; Marchis et al. 2006, 2014; Mueller et al. 2010). These low densities are consistent with the presence of ice(s) in the interior of these bodies.

**16.4 Irregular satellites of the giant planets**

16.4.1 Surface composition

The surface compositions of the irregular satellites of the giant planets have often been used to attempt to understand the origins of these bodies. Unfortunately, the information we have on their compositions is sparse. Visible through near-infrared photometry show generally featureless spectra with flat to slightly red slopes in the optical and slightly red slopes in the near-infrared (e.g., Grav et al. 2003; Grav & Holman 2004; Vilas et al. 2006). By analogy to asteroids, these characteristics often lead to these satellites to be referred to as C-type or P/D-type objects, with the implication of a common origin between the irregular satellites and these asteroids. The few visible and near-infrared spectra that have been collected so far for the brightest objects (e.g., Himalia, Phoebe, Triton and Nereid) seem to confirm this trend (e.g., Brown 2000). Himalia's spectrum in the 0.8-4 micron range is similar to those of C-types asteroids (Dumas et al. 1998; Brown 2000; Brown & Rhoden 2014). The same can be said for Phoebe, the difference being that in this case water ice bands are superimposed on a typical C-type spectrum (Brown 2000, Clark et al. 2005).

16.4.2 Bulk composition

There is only one irregular satellite for which the density has been determined precisely, namely Phoebe (~1.6 g/cm$^3$ ; Thomas 2010).

**16.5 Neptune Trojans**

16.5.1 Surface composition

There is very limited information concerning the physical properties of Neptune Trojans. Sheppard and Trujillo (2006) obtained optical colors for four Neptune Trojans and found that they are indistinguishable from each other. This is currently unlike the Kuiper Belt objects (KBOs), which show large color diversity from gray to some of the reddest objects in the solar system. They found that the Neptune Trojans have optical colors a little redder than the gray KBOs and consistent with those of i) the Jupiter Trojans, ii) the blue lobe of the centaur color distribution, iii) the irregular satellites of the gas giants, and possibly iv) the comets. In particular, the four Neptune Trojans do not have the extreme red colors shown by objects in the low-inclination ("cold"), high-perihelion classical Kuiper Belt (Sheppard and Trujillo 2006). Due to their faintness, no other measurements have been recorded (e.g., spectra, albedo); this implies that a large variety of surface compositions are still compatible with the observed colors.

16.5.2 Bulk composition

Currently, there are no density measurements that have been recorded for this population.

**16.6 KBOs**

16.6.1 Surface composition

Using all of the spectroscopic and photometric data as well as the current best understanding of the physics and chemistry of these bodies, Brown (2012) divided Kuiper belt surfaces into major groups, with differences being caused by size, formation location, and history:

1) Besides Haumea, all very large KBOs (D>~1500km) possess volatile-rich surfaces (mostly $N_2$ and $CH_4$; e.g., Barucci et al. 2005, Brown, Trujillo & Rabinowitz 2005, Cruikshank et al. 1997, Licandro et al. 2006, Schaller & Brown 2007a, Tegler et al. 2008). The presence of such volatiles on these objects' surfaces can be explained by a simple model of volatile loss and retention in the Kuiper belt (Schaller & Brown 2007b). In this model, all KBOs are assumed to start with typical cometary abundances but then volatile escape to space occurs at a rate determined by the surface temperature and gravity. Volatiles such as $N_2$ and $CH_4$ can thus be retained on objects, which are either massive enough or cold enough. The presence of pure water ice on Haumea's surface is best explained by a single giant impact earlier in the history of the solar system (e.g., Brown et al. 2007) that would have excavated Haumea's $CH_4$-rich crust as well as a significant portion of its icy mantle, which then became Haumea's satellites and collisional family.

2) The surface composition of KBOs in the D~1000-1500km size range can be seen as a transition between the $CH_4$-rich surfaces of the large KBOs and the water ice-rich surfaces of mid-sized (D~500-1000km) KBOs. Objects in this size range (e.g., Makemake, Quaoar, 2007 OR10) still contain $CH_4$ which appears to be the dominant molecule on their surface. Radiation products of methane-ice such as ethane are also present (e.g., Bennett et al. 2006, Merlin et al. 2010a).

3) Objects in the D~500-1000km size range possess water ice-rich surfaces (e.g., Barkume, Brown & Schaller 2008, Brown, Schaller & Fraser 2011, Barucci et al. 2011). In addition, ammonia is likely present on these objects' surfaces (e.g., Merlin et al. 2010a, Barucci et al. 2008a). Based on the sharp increase in water ice absorption with larger size for these objects, Brown (2012) hypothesize that on these largest objecs the presence of water ice is a remnant of past liquid flows on the surface.

4) Small objects (D<~500km), which are by far the most abundant KBOs, show a great variety of colors, from neutral to very red ones (e.g., Peixinho et al. 2012). Water ice appears to be relatively common on these objects' surfaces (e.g., Barkume, Brown & Schaller 2008). Strikingly, it appears that the dynamically stable cold classical KBOs possess uniform red colors whereas the remainder of the KBOs have no discernible pattern to their colors (e.g.,

Peixinho et al. 2008). Brown et al. (2011) propose a chemical and dynamical process to explain this color diversity. Specifically, they propose that :

- the objects formed in the inner part of the primordial belt (inside of ~20AU) retained only $H_2O$ and $CO_2$ as the major ice species on their surfaces. Irradiation of these species then caused dark neutrally colored surfaces to develop.
- the objects formed further in the disk (outside of ~20AU) retained $CH_3OH$, which has been shown to lead to brighter redder surfaces after irradiation.

As of today, most of the compositional constraints we have for TNOs concern the nature of the volatiles present at their surfaces. In particular, little is known about the composition of their refractory components and whether dust grains are present at their surfaces.

Of great interest, mid-infrared observations have been used to constrain the surface composition of the centaur Absolus and in particular the mineralogy of its silicates. To date, Absolus is the only 'KBO' [*Centaurs are former KBOs, which have been perturbed onto short-lived planet-crossing orbits (Brown 2012). They appear indistinguishable from the smallest KBOs whose spectra can be measured: they contain either no detectable absorption features, a small amount of absorption due to water ice, or (in one case) absorption due to water ice and methanol (see Barucci et al. 2008b and references therein)*] for which a mid-infrared spectrum with sufficient signal-to-noise has been recorded. Its spectrum is remarquably similar to the Jupiter Trojan ones (Barucci et al. 2008b; Emery et al 2006) suggesting the presence of crystalline anhydrous silicates (olivine-rich IDP-like composition) in centaurs and thus KBOs (Vernazza et al. 2012) as it is the case for comets.

16.6.2 Bulk composition

Density measurements of TNOs have revealed a large variety of values, ranging from ~0.5 $g/cm^3$ to ~3 $g/cm^3$ (Brown 2012). Remarkably, small objects (D<400 km) possess low densities (mostly lower than ~1.0 $g/cm^3$), whereas large objects (>1000 km) possess high densities (sometimes as high as ~3 $g/cm^3$) (Brown 2013). As of today, this extreme bulk compositional variability is not understood (see Brown 2012 for more details).

**16.7 Comets**

Comets are currently the most important source of knowledge regarding the nature and relative abundance of the volatiles that condensed in the outer solar system and that have been incorporated into the giant planet cores and satellites as well as in the KBOs. They also carry important constraints regarding the processess that occured in the protoplanetary disk prior to planetesimal formation.

Over the last 40 years, compositional surveys of both short period (SP) and long period (LP) comets have been performed. Based on the latter, we now have a clearer understanding of the global composition of comets (nature of their volatiles, dust composition) and of possible compositional differences between SPCs and LPCs.

16.7.1 Volatiles

More than 20 primary chemical species have now been detected in comets via spectroscopic surveys at IR and radio wavelengths (Bockelée-Morvan et al. 2004, Mumma & Charnley 2011): $H_2O$, $CO$, $CO_2$, $CH_4$, $C_2H_2$, $C_2H_6$, $CH_3OH$, $H_2CO$, $HOCH_2CH_2OH$, $HCOOH$,

HCOOCH$_3$, CH$_3$CHO, NH$_2$CHO, NH$_3$, HCN, HNCO, CH$_3$CN, HC$_3$N, H$_2$S, OCS, SO$_2$, H2CS and S$_2$. H$_2$O is the most abundant specie followed by CO$_2$. Note that two of the aforementioned species (CO, H$_2$CO) are in large part product species. Molecular abundances relative to water for the main species show the same dispersion and thus little if any systematic difference between LPCs and SPCs (Crovisier et al. 2009a,b; Mumma & Charnley 2011; A'Hearn et al. 2012).

16.7.2 Dust

Cometary dust has been studied via i) Earth based mid-infrared spectroscopic observations, ii) the *Stardust* samples from comet 81P/Wild 2, and iii) IDPs (interplanetary dust particles) collected in the Earth stratosphere.

Spectral data in the 8-13 micron region now exists for more than 20 comets (see Fig. 16.2). It appears that both LPCs and SPCs possess very similar spectral properties (Fig. 16.2), implying, on average, very similar dust compositions for both populations. Their spectra reveal the presence of anhydrous silicates (olivine, pyroxene), both in crystalline and amorphous phase (see Vernazza et al. 2015b and references therein). In both populations, there are objects enriched in crystalline olivine with respect to pyroxene (e.g., Hale-Bopp, Halley) whereas the remaining objects tend to have about as much crystalline pyroxene as olivine (Vernazza et al. 2015b). Overall, the spectral properties of LPCs and SPCs appear very similar to those of olivine-rich IDPs (Vernazza et al. 2015b) and to those of P- and D-type asteroids (and those of the Jupiter Trojans; e.g., Vernazza et al. 2015b), which suggests that both comets and P- and D-type asteroids are the likely source of olivine-rich IDPs.

Both the *Stardust* samples and cometary IDPs have confirmed the presence of amorphous and crystalline silicate material within cometary dust (e.g., Zolensky et al. 2008, Westphal et al. 2009). The *Stardust* samples further revealed the presence of CAIs and chondrules and that crystalline silicates are on average more abundant than amorphous ones (e.g., Zolensky et al. 2006, Brownlee 2006, Wooden 2008, Westphal et al. 2009). Alltogether, these observations imply that a large fraction of the cometary dust (crystalline silicates, CAIs, chondrules) formed in the 'warm' inner solar system and was subsequently transported outward to the 'cold' comet-forming zones via a to be identified mecanism (Ciesla 2007, Kelley and Wooden 2009).

16.7.3 Bulk composition

Up to now, there is only one comet for which the density has been determined with high precision (0.48±0.05 g/cm$^3$), namely 67P which has been visited closeby by Rosetta (Sierks et al. 2015). Such low density is compatible with the one that has been measured for small (D<500km) TNOs (see Brown 2013 and references therein).

**16.8 Decoding the solar system's past**

Current solar system formation models (e.g., Grand Tack and Nice models; Morbidelli et al. 2005, Bottke et al. 2006, Levison et al. 2009, Walsh et al. 2011) suggest that significant radial mixing of planetesimals has occurred during the early solar system mainly as a consequence of giant planet migrations. If all these models are correct, the asteroid belt and other populations of small bodies (e.g., KBOs, Trojans) should be condensed versions of the primordial solar system, containing objects that formed at large radial distances (> 10AU from the Sun) as well as planetesimals that formed close to the Sun (< 3AU from the Sun). In this context, constraining the surface composition of solar system small bodies and establishing the compositional link between the different classes of small bodies provides key constraints to these models and allows making progress on the following key questions: ''What was the compositional gradient in the solar system at the time of planetary formation?'', and ''What was the composition of the building blocks of the telluric planets and the giant planet cores?''

16.8.1 Constraints on current dynamical models (Grand Tack, Nice model)

As of today, it appears that the asteroid belt is the only population of small bodies that may effectively be a condensed version of the solar system (Fig 16.3A). Little compositional variety concerning the non-volatile phase (rocky component) is observed among the remaining populations of small bodies (Jupiter Trojans, KBOs, centaurs, comets) for which the rocky component appears to be similar to olivine-rich IDPs. There seems to be one exception in the outer solar system (>5AU), with the presence of C-type bodies along P- and D-types among the irregular satellites of Jupiter and Saturn.

In particular, there is currently no evidence of A, E, K, L, M, O, V-, or S-type bodies among Jupiter trojans (Marsset et al. 2014, DeMeo & Carry 2013), comets and KBOs (Brown 2012). This momentanly precludes significant outward migration of small bodies formed in the inner solar system (<4AU) as advocated by the Grand Tack model. On the other hand, significant inward migration of outer solar system small bodies as predicted by the Nice model is coherent with the compositional diversity of the asteroid belt.

16.8.2 Primordial architecture

16.8.2.1 Spatial distribution

One of the main goals in planetary science is to retrieve the initial compositional distribution of small bodies across the solar system (Fig. 16.3B). Ultimately, a proper understanding of the initial compositional gradient combined with the present day distribution will allow us to decipher the dynamical evolution of our solar system as a function of time. It will also allow us to characterize the sequence of events (e.g., dust migration) that occured in the protoplanetary solar disk before small bodies and planets formed. To achieve this goal, one needs to combine a considerable amount of information (see a first list below) into a coherent scheme.

In figure 16.3B, we provide a first guess of the initial compositional distribution using the following constraints (note: we used the book by Hutchison (2004) to retrieve informations related to meteorites):
- the compositional distribution in the asteroid belt and among outer solar system populations (see previous sections)
- the chondrule occurence (or chondrule fragments) in associated extraterrestrial materials (meteorites, IDPs, comet samples)
- the water content of small bodies inferred from their density and surface composition (see previous sections)
- the hydrothermal history of the associated extraterrestrial materials (meteorites, IDPs)
- the redox state of associated extraterrestrial materials (meteorites, IDPs)
- the locations of the snowline and methanol line
- the density range of the different classes of small bodies (e.g., Carry 2012)

- the affinitites between extraterrestrial materials based on oxygen isotopes ratio in particular [based on the latter, ECs are proposed by some researchers to be plausible building blocks of the Earth (e.g., Javoy et al. 2010), and OCs may be plausible building blocks of Mars (Burbine & O'Brien 2004)]

### 16.8.2.2 Temporal distribution

We also introduce the notion that all classes of small bodies did not necessarily form at the same time, as suggested by meteorites studies (e.g., Dauphas and Chaussidon 2011). The compositional distribution of small bodies across the solar system is usually interpreted as a consequence of the formation location whereas the role played by the time of formation on the resulting distribution is being overlooked. Here, we provide a first guess of the sequence of formation of the different classes of small bodies using the following constraints:
- the presence or absence of a remnant magnetic field in associated extraterrestrial materials (e.g., Weiss & Elkins-Tanton 2013)
- the peak temperature experiences by associated extraterrestrial materials in the parent body
- links between chondrites and achondrites based on isotopic ratios (aubrites and ECs)
- the paucity/abundance of objects in the different classes of small bodies ;  we consider that the classes that contain only a few objects may be the most ancient generations of bodies whereas the classes containing most of the objects may be the last generations of bodies to form (same trend as for human civilizations)
- the existence of metallic asteroids (cores of differentiated protoplanets ?) combined with the paucity of olivine-rich asteroids (mantles of differentiated protoplanets ?) may suggest that primordial protoplanets have been battered to bits (Burbine et al. 1996)
- the nature of mesosiderites which imply a violent collisional history during the early solar system that would be visible among the last generations of small bodies if it had occured later on
- the abundance of CAIs in associated extraterrestrial materials and small bodies (e.g., L-types;  see Sunshine et al. 2008)

Obviously, what we propose here are preliminary ideas rather than a definitive theory of "what precisely happened" during the early solar system; our main motivation is to define new routes for future research. Therefore, Figure 16.3b should not be taken too literally and we expect these ideas to evolve considerably over the next decades.

### 16.9 Conclusion: open questions

The general understanding of the compositions present among the different dynamical classes of solar system small bodies has greatly improved over the last decade, mainly because of the emergence of numerous high quality spectroscopic measurements in the near- and mid-infrared ranges but also because of several succesful space missions (Hayabusa, Dawn, Rosetta, etc..). We now realize that the compositional distribution of small bodies across the solar system holds many of the keys necessary to decode the solar system's dynamical past.

We list below a series of questions and/or measurements (the list is not meant to be exhausive) that we believe should be investigated/conducted in the future and that could help us get one step closer to deciphering the solar system history:

- We still need a compositional characterization - including mid-infrared spectroscopic measurements - to be performed for a statistically significant sample of outer solar system small bodies, to :

        - constrain their dust composition
        - search for the presence of A, E, K, L, M, O, V-, and S-type bodies in order to further test Grand Tack-like models

-What is the mechanism that produced the mineralogical gradient between BCG-, P- and D-types ? Future sample return missions (OSIRIS-Rex, Hayabusa 2) will help addressing this issue.

- What kind of volatiles/ices where intially present among C-complex asteroids? what kind of volatiles/ices are still present among C-complex asteroids? the same question can be asked for P- and D-types although comets have already partially answered the question.

- Which classes of small bodies were the main building blocks of Mercury and Venus? is it correct to assume that ECs and OCs were the main building blocks of respectively the Earth and Mars (e.g., Javoy et al. 2010, Burbine & O'Brien 2004) ?

- Where are the CI and Tagish Lake parent bodies and what do they represent (e.g., interior of large BCG- and/or P and D-types) ? The same question applies to hydrated IDPs.

- Are there many more R-, O-, A-, E, M-, K- and L-types at small sizes (D~5-20km) than at large ones (D>30km) ? This could potentially allow testing the idea that these types are more ancient and have been battered to bits (Burbine et al. 1996).

- Do C-complex asteroids and the regular satellites of Jupiter and Saturn share the same building blocks (as suggested by Fig. 16.3B)? Are their volatile isotopic ratios similar ?

- LPCs and SPCs share similar compositions. But are there, on average, statistically significant compositional differences between the two populations?

- Are D>100km bodies and even D>1000km present in the Oort cloud and if so what is the compositional distribution within this population?

- The formation location of chondrules seems to be localized (< 4A.U.). What are the implications for the nature of the chondrule formation process ?

- All solar system bodies apart from the telluric planets are predicted to have migrated at some point in their history following current dynamical models. Is there any reason why the telluric planets haven't?

## Acknowledgments

We wish to thank both referees for their constructive comments, which helped to improve the manuscript.

## 16.10 References

Abell, P. A., Vilas, F., Jarvis, K. S., *et al.* 2007. Mineralogical composition of (25143) Itokawa 1998 SF36 from visible and near-infrared reflectance spectroscopy: Evidence for partial melting. *Meteoritics & Planetary Science*, **42**, 2165-2177.

A'Hearn, M. F., Feaga, L. M., Keller, H. U., *et al.* 2012. Cometary Volatiles and the Origin of Comets. *The Astrophysical Journal*, **758**, article id. 29, 8 pp.

Barkume, K.M., Brown, M.E., Schaller, E.L. 2008. Near-Infrared Spectra of Centaurs and Kuiper Belt Objects. *Astron. J.*, **135**, 55-67.

Barucci, M. A., Dotto, E., Brucato J., *et al.* 2002. 10 Hygiea: ISO Infrared Observations. *Icarus*, **156**, 202–210.

Barucci, M.A., Cruikshank, D.P., Dotto, E., *et al.* 2005. Is Sedna another Triton? *Astron. Astrophys.*, **439**, L1-L4.

Barucci, M.A., Merlin, F., Guilbert, A., *et al.* 2008a. Surface composition and temperature of the TNO Orcus. *Astron. Astrophys.*, **479**, L13-L16.

Barucci, M.A., Brown, M.E., Emery, J.P., Merlin, F. 2008b. Composition and Surface Properties of Transneptunian Objects and Centaurs, In The Solar System Beyond Neptune, eds. MA Barucci, H Boehnhardt, DP Cruikshank, A Morbidelli. 143-160.

Barucci, M.A., Alvarez-Candal, A., Merlin, F., *et al.* 2011. New insights on ices in Centaur and Transneptunian populations. *Icarus*, **214**, 297-307.

Bell, J. F. 1988. A probable asteroidal parent body for the CV or CO chondrites (abstract). *Meteoritics*, **23**, 256–257.

Bennett, C.J., Jamieson, C.S., Osamura, Y., Kaiser, R.I. 2006. Laboratory Studies on the Irradiation of Methane in Interstellar, Cometary, and Solar System Ices. *Astrophys. J.*, **653**, 792-811.

Binzel, R. P., Xu, S., Bus, S. J., *et al.* 1993. Discovery of a Main-Belt Asteroid Resembling Ordinary Chondrite Meteorites. *Science*, **262**, 1541-1543.

Bockelée-Morvan, D., Crovisier, J., Mumma, M. J., Weaver, H. A. 2004. The composition of cometary volatiles. In Comets II, University of Arizona Press, Tucson, 391-423.

Bottke, W. F., Jr., Cellino, A., Paolicchi, P., Binzel, R. P. 2002. An Overview of the Asteroids: The Asteroids III Perspective. Asteroids III, W. F. Bottke Jr., A. Cellino, P. Paolicchi, and R. P. Binzel (eds), University of Arizona Press, Tucson, 3-15.

Bottke, W. F., Nesvorny, D., Grimm, R. E., Morbidelli, A., O'Brien, D. P. 2006. Iron meteorites as remnants of planetesimals formed in the terrestrial planet region. *Nature*, **439**, 821-824.

Bottke, W. F., Vokrouhlický, D., Minton, D., *et al.* 2012. An Archaean heavy bombardment from a destabilized extension of the asteroid belt. *Nature*, **485**, 78-81.

Bradley, J. P., Keller, L. P., Brownlee, D. E., & Thomas, K. L. 1996. Reflectance spectroscopy of interplanetary dust particles. *Meteoritics and Planetary Science*, **31**, 394–402.

Bradley, J. P., Keller, L. P., Snow, T. P., *et al.* 1999. An infrared spectral match between GEMS and interstellar grains. *Science*, **285**, 1716–1718.

Brown, M. E. 2000. Near-Infrared Spectroscopy of Centaurs and Irregular Satellites. *The Astronomical Journal*, **119**, 977-983.

Brown, M.E., Trujillo, C.A., Rabinowitz, D.L. 2005. Discovery of a Planetary-sized Object in the Scattered Kuiper Belt. *Astrophys. J. Lett.*, **635**, L97-L100.

Brown, M.E., Barkume, K.M., Ragozzine, D., Schaller, E.L. 2007. A collisional family of icy objects in the Kuiper belt. *Nature*, **446**, 294-296.

Brown, M. E., Schaller, E. L., Fraser, W. C. 2011. A Hypothesis for the Color Diversity of the Kuiper Belt. *The Astrophysical Journal Letters*, **739**, article id. L60, 5 pp.

Brown, M.E., Schaller, E.L., Fraser, W.C. 2012. Water Ice in the Kuiper Belt. *Astron. J.*, **143**, 7 pp.

Brown, M. E. 2012. The Compositions of Kuiper Belt Objects. *Annual Review of Earth and Planetary Sciences*, **40**, 467-494.

Brown, M. E. 2013. The Density of Mid-sized Kuiper Belt Object 2002 UX25 and the Formation of the Dwarf Planets. *The Astrophysical Journal Letters*, **778**, 5 pp.

Brown, M. E. & Rhoden, A. R. 2014. The 3 mum Spectrum of Jupiter's Irregular Satellite Himalia. *The Astrophysical Journal Letters*, **793**, 3 pp.

Brownlee, D., Tsou, P., Aléon, J., *et al.* 2006. Comet 81P/Wild 2 Under a Microscope. *Science*, **314**, 1711-1716.

Brunetto, R., Romano, F., Blanco, A., *et al.* 2006. Space weathering of silicates simulated by nanosecond pulse UV excimer laser. *Icarus*, **180**, 546–554.

Burbine, T. H., Gaffey, M. J., Bell, J. F. 1992. S-asteroids 387 Aquitania and 980 Anacostia - Possible fragments of the breakup of a spinel-bearing parent body with CO3/CV3 affinities. *Meteoritics*, **27**, 424-434.

Burbine, T. H., Meibom, A., Binzel, R. P. 1996. Mantle material in the main belt: Battered to bits? *Meteoritics & Planetary Science*, **31**, 607-620.

Burbine, T. H. 1998. Could G-class asteroids be the parent bodies of the CM chondrites? *Meteoritics & Planetary Science*, **33**, 253-258.

Burbine, T. H., Binzel, R. P., Bus, S. J., Clark, B. E. 2001. K asteroids and CO3/CV3 chondrites. *Meteoritics & Planetary Science*, **36**, 245-253.

Burbine, T. H., McCoy, T. J., Meibom, A., *et al.* 2002. Meteoritic Parent Bodies: Their Number and Identification. In Asteroids III. University of Arizona Press, Tucson, 653-667.

Burbine, T. H., O'Brien, K. M. 2004. Determining the possible building blocks of the Earth and Mars. *Meteoritics & Planetary Science*, **39**, 667-681.

Burbine, T. H., McCoy, T. J., Hinrichs, J. L., Lucey, P. G. 2006. Spectral properties of angrites. *Meteoritics & Planetary Science*, **41**, 1139-1145.

Burbine, T. H., Duffard, R., Buchanan, P. C., *et al.* 2011. Spectroscopy of O-Type Asteroids. *42nd LPSC*, **1608**, 2483.

Burbine, T. H. 2014. Asteroids. Planets, Asteriods, Comets and The Solar System, Volume 2 of Treatise on Geochemistry (Second Edition). Edited by Andrew M. Davis. Elsevier, 365-415.

Bus, S. J. 1999. Compositional Structure in the Asteroid Belt: Results of a Spectroscopic Survey. Ph.D. thesis, Massachusetts Inst. Technol.

Bus, S. J. & Binzel, R. P. 2002. Phase II of the Small Main-Belt Asteroid Spectroscopic Survey: a feature-based taxonomy. *Icarus*, **158**, 146–177.

Campins, H., & Ryan, E. V. 1989. The identification of crystalline olivine in cometary silicates. *The Astrophysical Journal*, **341**, 1059–1066.

Campins, H., Ziffer, J., Licandro *et al.* 2006. Nuclear Spectra of Comet 162P/Siding Spring (2004 TU12). *The Astronomical Journal*, **132**, 1346–1353.

Campins, H., Licandro, J., Pinilla-Alonso, N. *et al.* 2007. Nuclear Spectra of Comet 28P Neujmin 1. *The Astronomical Journal*, **134**, 1626–1633.

Campins, H. *et al.* 2010. Water ice and organics on the surface of the asteroid 24 Themis. *Nature*, **464**, 1320–1321.

Capaccioni, F., Coradini, A., Filacchione, G., *et al.* 2015. The organic-rich surface of comet 67P/Churyumov-Gerasimenko as seen by VIRTIS/Rosetta. *Science*, **347**, 0628.

Carry, B. 2012. Density of asteroids. *Planet. Space Sci.*, **73**, 98–118.

Chapman C.R., Salisbury J.W. 1973. Comparisons of meteorite and asteroid spectral reflectivities. *Icarus*, **19**, 507–22.

Chapman, C.R. 2004. Space weathering of asteroid surfaces. *Annu. Rev. Earth Planet. Sci.*, **32**, 539–567.

Chiang, E. & Youdin, A. N. 2010. Forming Planetesimals in Solar and Extrasolar Nebulae. *Annual Review of Earth and Planetary Sciences*, **38**, 493-522.

Ciesla, F. J. 2007. Outward Transport of High-Temperature Materials Around the Midplane of the Solar Nebula. *Science*, **318**, 613-615.

Ciesla, F. J., Davison, T. M., Collins, G. S., O'Brien, D. P. 2013. Thermal consequences of impacts in the early solar system. *Meteoritics & Planetary Science*, **48**, 2559-2576.

Clark, B.E., Bus, S.J., Rivkin, A.S., *et al.* 2004a. Spectroscopy of X-type asteroids. *Astronomy Journal* **128**, 3070–3081.

Clark, B.E., Bus, S.J., Rivkin, A.S., *et al.* 2004b. E-type asteroid spectroscopy and compositional modeling. *Journal of Geophysical Research*, **109**, E02001.

Clark, B.E., Ockert-Bell, M.E., Cloutis, E.A., *et al.* 2009. Spectroscopy of K-complex asteroids: Parent bodies of carbonaceous meteorites? *Icarus*, **202**, 119–133.

Clark, R. N., Brown, R. H., Jaumann, R., *et al.* 2005. Compositional maps of Saturn's moon Phoebe from imaging spectroscopy. *Nature*, **435**, 66-69.

Cloutis, E. A., Gaffey, M. J., Smith, D. G. W., and Lambert, R. St. J. 1990. Reflectance spectra of "featureless" materials and the surface mineralogies of M- and E-class asteroids. *J. Geophys. Res.*, **95**, 281–293.

Cloutis, E. A., Binzel, R. P., Burbine, T. H., *et al.* 2006. Asteroid 3628 Boznemcová: Covered with angrite-like basalts? *Meteoritics & Planetary Science*, **41**, 1147-1161.

Cohen, M., Witteborn, F. C., Roush, T., *et al.* 1998. Spectral Irradiance Calibration in the Infrared. VIII. 5-14 Micron Spectroscopy of the Asteroids Ceres, Vesta, and Pallas. *The Astronomical Journal*, **115**, 1671–1679.

Consolmagno, G. J. and Drake, M. J. 1977. Composition and evolution of the eucrite parent body: Evidence from rare earth elements. *Geochim. Cosmochim. Acta.*, **41**, 1271–1282.

Consolmagno, G., Britt, D., Macke, R. 2008. The significance of meteorite density and porosity. *Chemie der Erde*, **68**, 1-29.

Crovisier, J., Biver, N., Bockelée-Morvan, D., Colom, P. 2009a. Radio observations of Jupiter-family comets. *Planetary and Space Science*, **57**, 1162-1174.

Crovisier, J., Biver, N., Bockelée-Morvan, D., *et al.* 2009b. The Chemical Diversity of Comets: Synergies Between Space Exploration and Ground-based Radio Observations. *Earth, Moon, and Planets,* **105**, 267-272.

Cruikshank, D. P., and Hartmann, W. K. 1984. The meteorite-asteroid connection: Two olivine-rich asteroids. *Science*, **223**, 281–283.

Cruikshank, D.P., Roush, T.L., Moore, J.M., *et al.* 1997. The Surfaces of Pluto and Charon, In Pluto and Charon, ed. Stern, S. A. & Tholen, D. J. 221-+.

De León, J., Licandro, J., Serra-Ricart, M., *et al.* 2010. Observations, compositional, and physical characterization of near-Earth and Mars-crosser asteroids from a spectroscopic survey. *Astronomy and Astrophysics*, **517**, id.A23, 25 pp.

De Luise, F., Dotto, E., Fornasier, S., *et al.* 2010. A peculiar family of Jupiter Trojans: The Eurybates. *Icarus*, **209**, 586-590.

DeMeo, F. E., Binzel, R. P., Slivan, S. M., Bus, S. J. 2009. An extension of the Bus asteroid taxonomy into the near-infrared. *Icarus*, **202**, 160-180.

DeMeo, F. E. & Carry, B. 2013. The taxonomic distribution of asteroids from multi-filter all sky photometric surveys. *Icarus*, **226**, 723–741.

DeMeo, F. E. & Carry, B. 2014. Solar System evolution from compositional mapping of the asteroid belt. *Nature*, **505**, 629-634.

Dumas, C., Owen, T., Barucci, M. A. 1998. Near-Infrared Spectroscopy of Low-Albedo Surfaces of the Solar System: Search for the Spectral Signature of Dark Material. *Icarus*, **133**, 221-232.

Dunn, T. L., Burbine, T. H., Bottke, W. F., Clark, J. P. 2013. Mineralogies and source regions of near-Earth asteroids. *Icarus*, **222**, 273-282.

Emery, J. P. & Brown, R. H. 2003. Constraints on the surface composition of Trojan asteroids from near-infrared (0.8-4.0 mum) spectroscopy. *Icarus*, **164**, 104-121.

Emery, J.P., Cruikshank, D.P., van Cleve, J., 2006. Thermal emission spectroscopy (5.2-38 mum) of three Trojan asteroids with the Spitzer Space Telescope: Detection of fine-grained silicates. *Icarus*, **182**, 496–512.

Emery, J. P., Burr, D. M., Cruikshank, D. P. 2011. Near-infrared Spectroscopy of Trojan Asteroids: Evidence for Two Compositional Groups. *The Astronomical Journal*, **141**, 18 pp.

Emery, J. P., *et al.* 2015. The Complex History of Trojan Asteroids. In Asteroids IV. University of Arizona Press.

Gaffey, M. J., Bell, J. F., and Cruikshank, D. P. 1989. Reflectance spectroscopy and asteroid surface mineralogy. In Asteroids II. Tucson, AZ: University of Arizona Press, 98–127.

Gaffey, M.J., Bell, J.F., Brown, R.H., *et al.* 1993. Mineralogic variations within the S-type asteroid class. *Icarus*, **106**, 573–602.

Gomes, R., Levison, H.F., Tsiganis, K., Morbidelli, A. 2005. Origin of the cataclysmic Late Heavy Bombardment period of the terrestrial planets. *Nature*, **435**, 466–469.

Gradie, J., and Tedesco, E. 1982. Compositional structure of the asteroid belt. *Science*, **216**, 1405-1407.

Grav, T., Holman, M. J., Gladman, B. J., Aksnes, K. 2003. Photometric survey of the irregular satellites. *Icarus*, **166**, 33-45.

Grav, T. & Holman, M. J. 2004. Near-Infrared Photometry of the Irregular Satellites of Jupiter and Saturn. *The Astrophysical Journal*, **605**, L141-L144.

Hanner, M. S., Lynch, D. K., & Russell, R. W. 1994. The 8-13 micron spectra of comets and the composition of silicate grains. *The Astrophysical Journal*, **425**, 274–285.

Hanner, M. S., Gehrz, R. D., Harker, D. E., *et al.* 1997. Thermal Emission From The Dust Coma Of Comet Hale-Bopp And The Composition Of The Silicate Grains. *Earth Moon and Planets*, **79**, 247–264.

Hargrove, K. D., Kelley, M. S., Campins, H. *et al.* 2012. Asteroids (65) Cybele, (107) Camilla and (121) Hermione: Infrared spectral diversity among the Cybeles. *Icarus*, **221**, 453-455.

Harker, D. E., Woodward, C. E., Kelley, M. S., *et al.* 2011. Mid-infrared Spectrophotometric Observations of Fragments B and C of Comet 73P/Schwassmann-Wachmann 3. *The Astronomical Journal*, **141**, 26.

Hayward, T. L., Hanner, M. S., & Sekanina, Z. 2000. Thermal Infrared Imaging and Spectroscopy of Comet Hale-Bopp (C/1995 O1). *The Astrophysical Journal*, **538**, 428–455.

Hiroi, T., Pieters, C. M., Zolensky, M. E., Lipschutz, M. E. 1993. Evidence of thermal metamorphism on the C, G, B, and F asteroids. *Science*, **261**, 1016-1018.

Hiroi, T., Sasaki, S. 2001. Importance of space weathering simulation products in compositional modeling of asteroids: 349 Dembowska and 446 Aeternitas as examples. *Meteoritics & Planetary Science*, **36**, 1587-1596.

Hsieh, H. H., Jewitt, D. A. 2006. Population of Comets in the Main Asteroid Belt. *Science*, **312**, 561-563.

Hutchison, R. 2004. Meteorites: A Petrologic, Chemical and Isotopic Synthesis (Cambridge Univ. Press).

Javoy, M., Kaminski, E., Guyot, F., *et al.* 2010. The chemical composition of the Earth: Enstatite chondrite models. *Earth and Planetary Science Letters*, **293**, 259-268.

Jewitt, D. (2012). The active asteroids. *Astron. J.*, **143**, 66.

Johansen, A., Oishi, J. S., Mac, Low M.-M., *et al.* 2007. A. Rapid planetesimal formation in turbulent circumstellar disks. *Nature*, **448**, 1022-1025.

Johansen, A., Klahr, H., Henning, Th. 2011. High-resolution simulations of planetesimal formation in turbulent protoplanetary discs. *Astronomy & Astrophysics*, **529**, id.A62.

Johansen, A., Youdin, A. N., Lithwick, Y. 2012. Adding particle collisions to the formation of asteroids and Kuiper belt objects via streaming instabilities. *Astronomy & Astrophysics*, **537**, id.A125.

Kelley, M. S. and Gaffey, M. J. 2002. High-albedo asteroid 434 Hungaria: Spectrum, composition and genetic connections. *Meteoritics and Planetary Science*, **37**, 1815–1827.

Kelley, M. S. & Wooden, D. H. 2009. The composition of dust in Jupiter-family comets inferred from infrared spectroscopy. *Planetary and Space Science*, **57**, 1133-1145.

Küppers, M., O'Rourke, L., Bockelée-Morvan, D., Zakharov, V., Lee, S., *et al.* 2014. Localized sources of water vapour on the dwarf planet (1) Ceres. *Nature*, **505**, 525-527.

Levison, H. *et al.* 2009. Contamination of the asteroid belt by primordial trans-Neptunian objects. *Nature*, **460**, 364–366.

Licandro, J., Pinilla-Alonso, N., Pedani, M., *et al.* 2006. The methane ice rich surface of large TNO 2005 FY9: a Pluto-twin in the trans-neptunian belt? *Astron. Astrophys.*, **445**, L35-L38.

Licandro, J., Hargrove, K., Kelley, M., *et al.* 2012. 5-14 μm Spitzer spectra of Themis family asteroids. *Astronomy and Astrophysics*, **537**, 73.

Lisse, C. M., VanCleve, J., Adams, A. C., *et al.* 2006. Spitzer Spectral Observations of the Deep Impact Ejecta. *Science*, **313**, 635–640.

Lynch, D. K., Russell, R. W., & Sitko, M. L. 2002. 3- to 14-μm Spectroscopy of Comet C/1999 T1 (McNaught-Hartley). *Icarus*, **159**, 234–238.

Marchi, S., Brunetto, R., Magrin, S., *et al.* 2005. Space weathering of near-Earth and main belt silicate-rich asteroids: Observations and ion irradiation experiments. *Astron. Astrophys.*, **443**, 769–775.

Marchis, F., Hestroffer, D., Descamps, P., *et al.* 2006. A low density of 0.8gcm$^{-3}$ for the Trojan binary asteroid 617 Patroclus. *Nature*, **439**, 565-567.

Marchis, F., Enriquez, J. E., Emery, J. P., Mueller, M., Baek, M., *et al.* 2012. Multiple asteroid systems: Dimensions and thermal properties from Spitzer Space Telescope and ground-based observations. *Icarus*, **221**, 1130-1161.

Marchis, F., Durech, J., Castillo-Rogez, J., *et al.* 2014. The Puzzling Mutual Orbit of the Binary Trojan Asteroid (624) Hektor. *The Astrophysical Journal Letters*, **783**, 6 pp.

Marsset, M., Vernazza, P., Gourgeot, F., *et al.* 2014. Similar origin for low- and high-albedo Jovian Trojans and Hilda asteroids? *Astronomy & Astrophysics*, **568**, id.L7, 4 pp.

Merlin, F., Barucci, M.A., de Bergh, C., *et al.* 2010a. Chemical and physical properties of the variegated Pluto and Charon surfaces. *Icarus*, **210**, 930-943.

Merouane, S., Djouadi, Z., & Le Sergeant d'Hendecourt, L. 2014. Relations between Aliphatics and Silicate Components in 12 Stratospheric Particles Deduced from Vibrational Spectroscopy. *The Astrophysical Journal*, **780**, 174.

Morbidelli, A., Levison, H.F., Tsiganis, K., Gomes, R. 2005. Chaotic capture of Jupiter's Trojan asteroids in the early Solar System. *Nature*, **435**, 462–465.

Morbidelli, A., Levison, H. F., Gomes, R. 2008. The Dynamical Structure of the Kuiper Belt and Its Primordial Origin. In The Solar System Beyond Neptune, University of Arizona Press, 275-292.

Mueller, M., Marchis, F., Emery, J.P., *et al.* 2010. Eclipsing binary Trojan asteroid Patroclus: Thermal inertia from Spitzer observations. *Icarus*, **205**, 505–515.

Mumma, M. J., and Charnley, S. B. 2011. The Chemical Composition of Comets - Emerging Taxonomies and Natal Heritage. *Annual Review of Astronomy and Astrophysics*, **49**, 471-524.

Nakamura, T., Noguchi, T., Tanaka, M., *et al.* 2011. Itokawa dust particles: A direct link between S-type asteroids and ordinary chondrites. *Science*, **333**, 1113-1116.

Nakashima, D., Kita, N. T., Ushikubo, T., *et al.* 2014. Oxygen three-isotope ratios of silicate particles returned from asteroid Itokawa by the Hayabusa spacecraft: A strong link with equilibrated LL chondrites. *Earth and Planetary Science Letters*, **379**, 127-136.

Peixinho, N., Lacerda, P., Jewitt, D. 2008. Color-Inclination Relation of the Classical Kuiper Belt Objects. *The Astronomical Journal*, **136**, 1837-1845.

Peixinho, N., Delsanti, A., Guilbert-Lepoutre, A., *et al.* 2012. The bimodal colors of Centaurs and small Kuiper belt objects. *Astronomy & Astrophysics*, **546**, id.A86, 12 pp.

Rivkin, A. S. & Emery, J. P. 2010. Detection of ice and organics on an asteroidal surface. *Nature*, **464**, 1322–1323.

Rivkin, A. S. 2012. The fraction of hydrated C-complex asteroids in the asteroid belt from SDSS data. *Icarus*, **221**, 744-752.

Russell, C. T., Raymond, C. A., Coradini, A., *et al.* 2012. Dawn at Vesta: Testing the Protoplanetary Paradigm. *Science*, **336**, 684-686.

Santos-Sanz, P., Lellouch, E., Fornasier, S., *et al.* 2012. "TNOs are Cool": A survey of the trans-Neptunian region. IV. Size/albedo characterization of 15 scattered disk and detached objects observed with Herschel-PACS. *Astronomy & Astrophysics*, **541**, id.A92, 18 pp.

Sasaki, S., Nakamura, K., Hamabe, Y., *et al.* 2001. Production of iron nanoparticles by laser irradiation in a simulation of lunar-like space weathering. *Nature*, **410**, 555–557.

Schaller, E.L. & Brown, M.E. 2007a. Detection of Methane on Kuiper Belt Object (50000) Quaoar. *Astrophys. J. Lett.*, **670**, L49-L51.

Schaller, E.L. & Brown, M.E. 2007b. Volatile Loss and Retention on Kuiper Belt Objects. *Astrophys. J. Lett.*, **659**, L61-L64.

Shepard, M. K., Taylor, P. A., Nolan, M. C., *et al.* 2015. A radar survey of M- and X-class asteroids. III. Insights into their composition, hydration state, & structure. *Icarus*, **245**, 38-55.

Sheppard, S. S. & Trujillo, C. A. 2006. A Thick Cloud of Neptune Trojans and Their Colors. *Science*, **313**, 511-514.

Sierks, H., Lamy, P., Barbieri, C., *et al.* 2011. Images of Asteroid 21 Lutetia: A Remnant Planetesimal from the Early Solar System. *Science*, **334**, 487-490.

Sierks, H., Barbieri, C., Lamy, P., *et al.* 2015. On the nucleus structure and activity of comet 67P/Churyumov-Gerasimenko. *Science*, **347**, article id. aaa1044.

Sitko, M. L., Lynch, D. K., Russell, R. W., & Hanner, M. S. 2004. 3-14 Micron Spectroscopy of Comets C/2002 O4 (Honig), C/2002 V1 (NEAT), C/2002 X5 (Kudo-Fujikawa), C/2002 Y1 (Juels-Holvorcem), and 69P/Taylor and the Relationships among Grain Temperature, Silicate Band Strength, and Structure among Comet Families. *The Astrophysical Journal*, **612**, 576–587.

Strazzulla, G., Dotto, E., Binzel, R., *et al.* 2005. Spectral alteration of the Meteorite Epinal (H5) induced by heavy ion irradiation: A simulation of space weathering effects on near-Earth asteroids. *Icarus*, **174**, 31–35.

Sunshine, J. M., Bus, S. J., Corrigan, C. M., *et al.* 2007. Olivine-dominated asteroids and meteorites: Distinguishing nebular and igneous histories. *Meteoritics & Planetary Science*, **42**, 155-170.

Sunshine, J. M., Connolly, H. C., McCoy, T. J., *et al.* 2008. Ancient Asteroids Enriched in Refractory Inclusions. *Science*, **320**, 514-516.

Takahashi, J., Itoh, Y., & Takahashi, S. 2011. Mid-Infrared Spectroscopy of 11 Main-Belt Asteroids. *Publications of the Astronomical Society of Japan*, **63**, 499–.

Takir, D. & Emery, J. P. 2012. Outer Main Belt asteroids: Identification and distribution of four 3-mum spectral groups. *Icarus*, **219**, 641-654.

Tegler, S.C., Grundy, W.M., Vilas, F., *et al.* 2008. Evidence of N2-ice on the surface of the icy dwarf Planet 136472 (2005 FY9). *Icarus*, **195**, 844-850.

Thomas, P. C. 2010. Sizes, shapes, and derived properties of the saturnian satellites after the Cassini nominal mission. *Icarus*, **208**, 395-401.

Tsiganis, K., Gomes, R., Morbidelli, A., Levison, H.F. 2005. Origin of the orbital architecture of the giant planets of the Solar System. *Nature*, **435**, 459–461.

Tsuchiyama, A., Uesugi, M., Uesugi, K., *et al.* 2014. Three-dimensional microstructure of samples recovered from asteroid 25143 Itokawa: Comparison with LL5 and LL6 chondrite particles. *Meteoritics and Planetary Science*, **49**, 172-187.

Vernazza, P., Binzel, R. P., Thomas, C. A., *et al.* 2008. Compositional differences between meteorites and near-Earth asteroids. *Nature*, **454**, 858-860.

Vernazza, P., Binzel, R. P., Rossi, A., Fulchignoni, M. & Birlan, M. 2009a. Solar wind as the origin of rapid weathering of asteroid surfaces. *Nature*, **458**, 993–995.

Vernazza, P., Brunetto, R., Binzel, *et al.* 2009b. Plausible parent bodies for enstatite chondrites and mesosiderites: Implications for Lutetia's fly-by. *Icarus*, **202**, 477-486.

Vernazza, P., Lamy, P., *et al.* 2011. Asteroid (21) Lutetia as a remnant of Earth's precursor planetesimals. *Icarus*, **216**, 650-659.

Vernazza, P., Delbo, M., King, P. L., *et al.* 2012. High surface porosity as the origin of emissivity features in asteroid spectra. *Icarus*, **221**, 1162-1172.

Vernazza, P., Fulvio, D., Brunetto, R., *et al.* 2013. Paucity of Tagish Lake-like parent bodies in the Asteroid Belt and among Jupiter Trojans. *Icarus*, **225**, 517-525.

Vernazza, P., Zanda, B., Binzel, R. P., *et al.* 2014. Multiple and Fast: The Accretion of Ordinary Chondrite Parent Bodies. *The Astrophysical Journal*, **791**, 22 pp.

Vernazza, P., Zanda, B., Nakamura, T., *et al.* 2015a. The Formation and Evolution of Ordinary Chondrite Parent Bodies. In Asteroids IV. University of Arizona Press.

Vernazza, P., Marsset, B., Beck, P. *et al.* 2015b. Interplanetary Dust Particles as Samples of Icy Asteroids. *The Astrophysical Journal*, **806**, 10 pp.

Vilas, F., Lederer, S. M., Gill, S. L., *et al.* 2006. Aqueous alteration affecting the irregular outer planets satellites: Evidence from spectral reflectance. *Icarus*, **180**, 453-463.

Walsh, K. J., Morbidelli, A., Raymond, S. N., O'Brien, D. P. & Mandell, A. M. 2011. A low mass for Mars from Jupiter's early gas-driven migration. *Nature*, **475**, 206–209.

Weiss, B. P. & Elkins-Tanton, L. T. 2013. Differentiated Planetesimals and the Parent Bodies of Chondrites. *Annual Review of Earth and Planetary Sciences*, **41**, 529-560.

Westphal, A. J., Fakra, S. C., Gainsforth, Z., *et al.* 2009. Mixing Fraction of Inner Solar System Material in Comet 81P/Wild2. *The Astrophysical Journal*, **694**, 18-28.

Wooden, D. H., Woodward, C. E., & Harker, D. E. 2004. Discovery of Crystalline Silicates in Comet C/2001 Q4 (NEAT). *The Astrophysical Journal Letters*, **612**, L77–L80.

Wooden, D. H. 2008. Cometary Refractory Grains: Interstellar and Nebular Sources. *Space Science Reviews*, **138**, 75-108.

Woodward, C. E., Jones, T. J., Brown, B., *et al.* 2011. Dust in Comet C/2007 N3 (Lulin). *The Astronomical Journal*, **141**, 181.

Yang, B., Lucey, P., Glotch, T. 2013. Are large Trojan asteroids salty? An observational, theoretical, and experimental study. *Icarus*, **223**, 359-366.

Youdin, A. N. & Goodman, J. 2005. Streaming instabilities in protoplanetary disks. *The Astrophysical Journal*, **620**, 459-469.

Youdin, A. N. 2011. On the formation of planetesimals via secular gravitational instabilities with Turbulent Stirring. *The Astrophysical Journal*, **731**, article id. 99.

Yurimoto, H., Abe, K., Abe, M., *et al.* 2011. Oxygen isotopic compositions of asteroidal materials returned from Itokawa by the Hayabusa mission. *Science*, **333**, 1116-1119.

Zellner, B. 1975. 44 Nysa: An iron-depleted asteroid. *The Astrophysical Journal*, **198**, L45–L47.

Zellner, B., Leake, M., Williams, J.G., and Morrison, D. 1977. The E asteroids and the origin of the enstatite achondrites. *Geochimica et Cosmochimica Acta*, **41**, 1759–1767.

Zolensky, M. E., Zega, T. J., Yano, H., *et al.* 2006. Mineralogy and Petrology of Comet 81P/Wild 2 Nucleus Samples. *Science*, **314**, 1735-1739.

Zolensky, M., Nakamura-Messenger, K., Rietmeijer, F., *et al.* 2008. Comparing Wild 2 particles to chondrites and IDPs. *Meteoritics & Planetary Science*, **43**, 261-272.

**Acknowledgments**

We wish to thank both referees for their constructive comments which helped improving the manuscript.

**Figures**

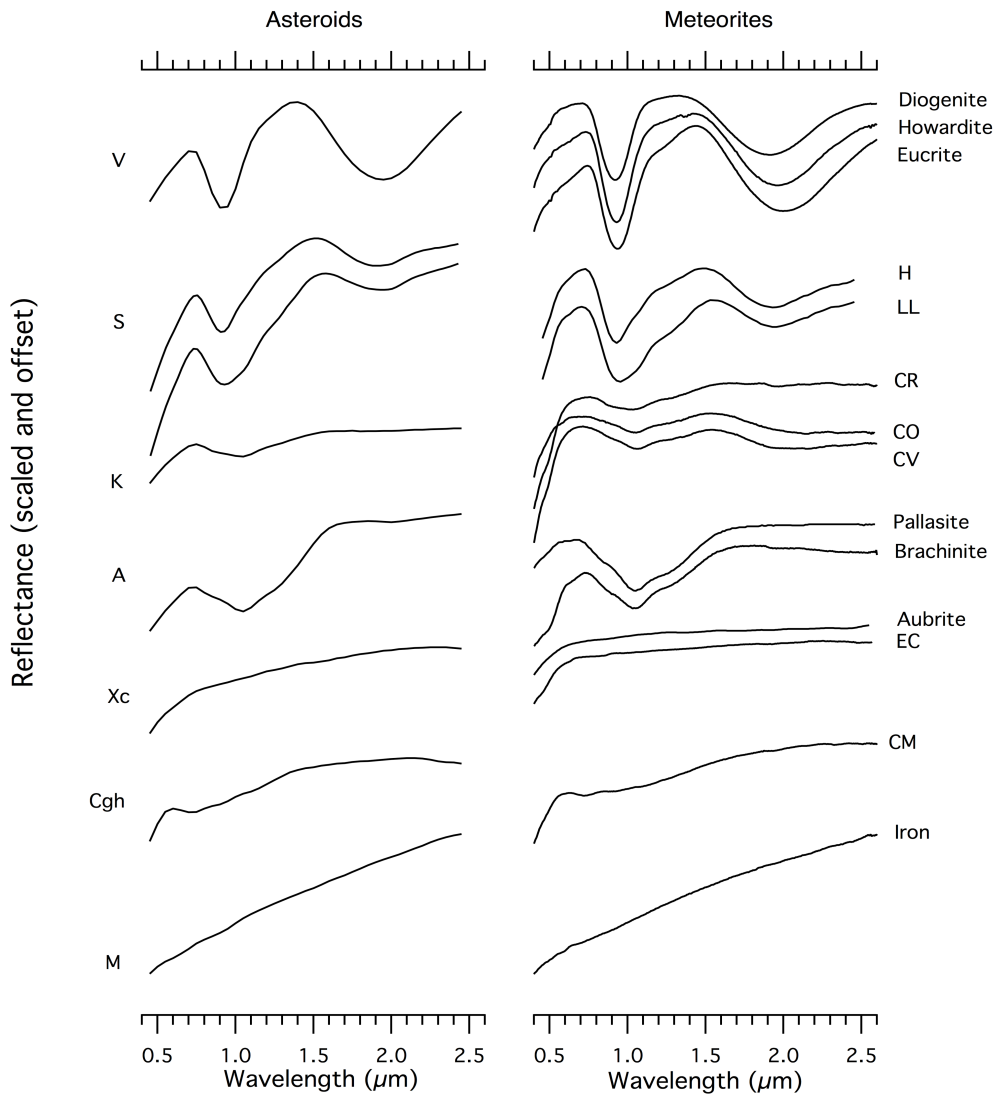

Figure 16.1: Visible and near-infrared spectra of several main-belt asteroids classes and their associated meteorites. The V, K, A, Xc, and Cgh asteroid spectra are endmembers from DeMeo et al., 2009. The M-type spectrum is an average of the high albedo (pv>0.1) X-type sectra from DeMeo et al. (2009). The two S-type spectra are averages of H-like (top) and LL-like (bottom) S-type asteroid spectra taken from Vernazza et al. (2014). The diogenite and eucrite spectra are averages of all available spectra in the RELAB database for these meteorite families. The howardite spectrum is that of EET 87503 (25-25 μm size fraction). The H, LL, CR, CO, CV, EC, CM and iron meteorite spectra are group averages calculated from the RELAB database. The pallasite spectrum is that of Thiel Mountain, digitized from Sunshine et al. 2007. The brachinite spectrum if that of Brachina (RELAB) while the aubrite spectrum is that of Mayo Belwa (RELAB). Spectra were normalozed to unity at 0.55 micron and offset vertically for clarity.

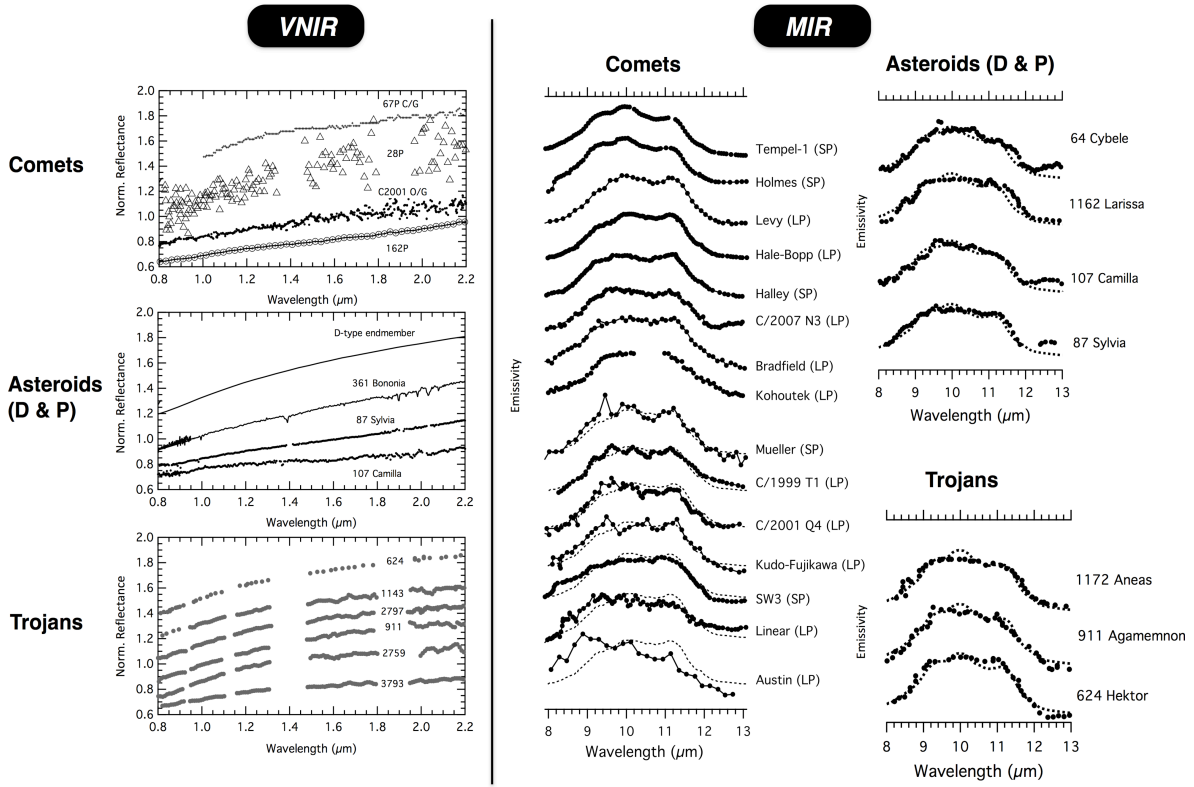

Figure 16.2: Observational evidences for the continuum in composition between comets, asteroids, centaurs and KBOs. We indicated hereafter, the source of the spectra. For the visible and near infrared (VNIR): the 67P spectrum was digitized from Capaccioni et al. (2015), that of C2001 O/G from Abell et al. (2006), and those of 28 P and 162 P from Campins et al. (2006,2007). For P- and D-type asteroids, the D-type endmember is from DeMeo et al. (2009) while the spectrum of Bononia is from Takir and Emery (2012). The VNIR spectra of Jupiter Trojans are from Emery and Brown (2003). For the mid-infrared (MIR) range, the comet spectra are from Campins & Ryan (1989), Hanner et al. (1994, 1997), Harker et al. (2011); Hayward et al. (2000), Lisse et al. (2006); Lynch et al. (2002), Sitko et al. (2004), Wooden et al. (2004); Woodward et al. (2011), the Jupiter Trojans' spectra are from Emery et al. (2006) while those of P- and D-type asteroids are from Marchis et al. (2012).

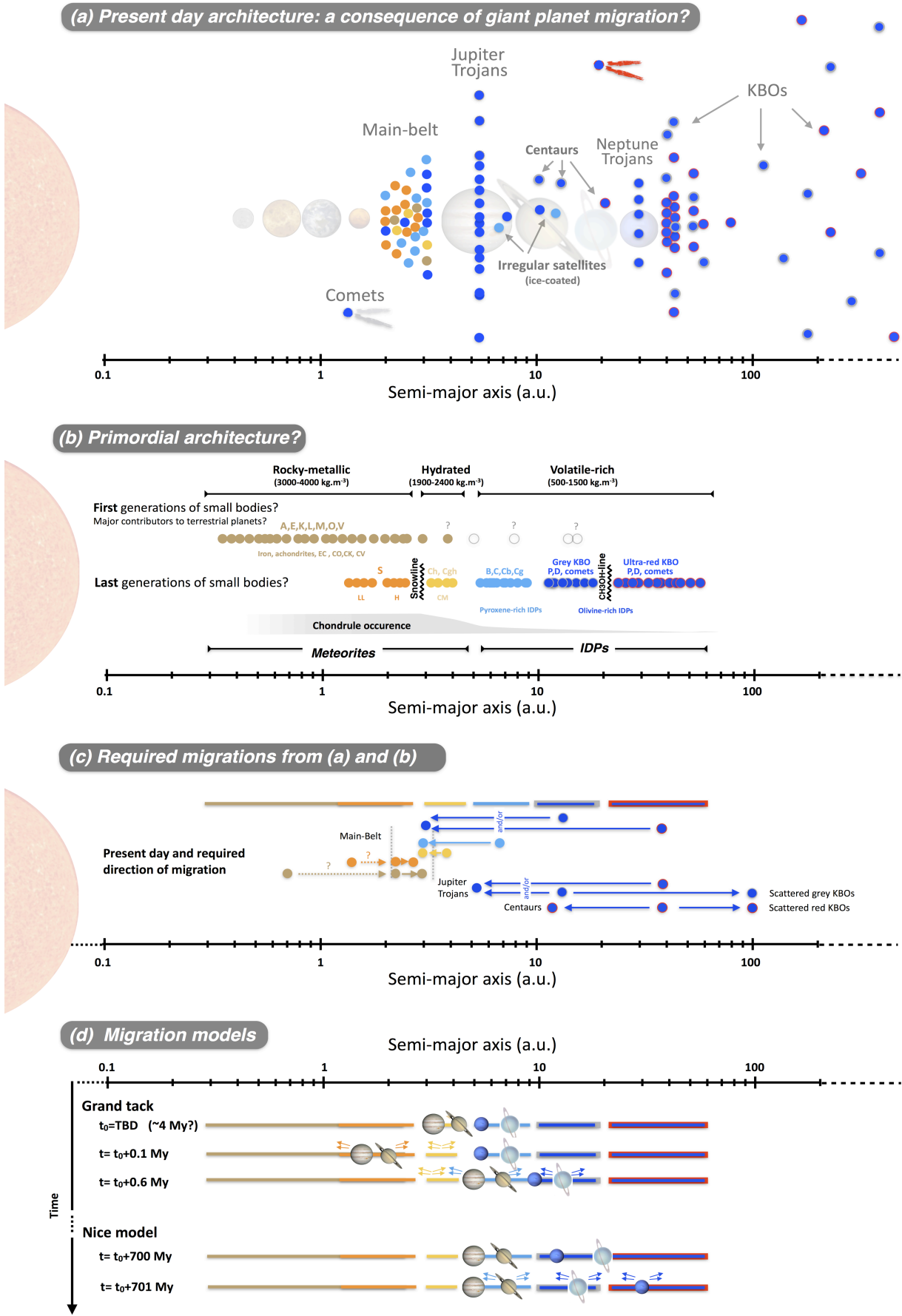

Figure 16.3: This figure describes the present-day architecture of small bodies (a) as well as an attempt to retrieve the primordial compositional gradient (b). From there, it is possible to evaluate the required migrations to achieve the present day structure (c), which can be confronted to migrations models (d).

**(a)** This panel presents the actual populations of solar system small bodies. They have been segmented into 5 major groups. First, a group containing A,E,K,M,L,M,O and V- types, whose associated meteorites present fingerprints of differentiation or heating (maroon). The four other groups are the S-types (orange), the Ch, Cgh (yellow), the B,C,Cb,Cg-types (light blue) and the P,D-types (dark blue). These four groups are identified from the combination of VNIR and MIR data and meteorite/IDP associations, as well as bulk densities. The display of these five colors among main-belt and Jupiter Trojan asteroids are in rough agreement with the radial and inclination distribution of the groups. The symbols of KBOs are blue (Trojan-like) since objects originating from this population (centaurs, Jupiter family comets) have Trojan-like spectra in the VNIR and MIR ranges. Future observations in the MIR will be required to confirm that KBOs have comet-like dust. The bimodal color distribution among KBOs is emphasized by red and grey aureoles, and their distribution is in rough agreement with observations. A bimodal color distribution is also observed among centaurs. Finally, the VNIR spectra of irregular satellites reveal that both B,C,Cb,Cg-like and P,D-like objects are present within these populations.

**(b)** This panel is a preliminary (raw) sketch of the primordial architecture of solar system small bodies. This diagram separates groups of bodies that likely accreted at different epochs, but not necessarily. Future research will need to clarify this hypothesis. For the A,E,K,L,M,O,V groups, associated meteorites usually reveal fast heating and eventually differentiation, an indication that they were among the first generations of planetesimals. Isotopic affinities (i.e. EC/aubrite with the Earth) open the possibility that at least some of these objects were formed in the terrestrial planet region. The presence of aqueous alteration fingerprints in some of them (CR,CV) suggest that they could have formed beyond the snowline. Their depleted nature with regard to other types of small bodies also points toward an early formation and subsequent depletion, possibly through terrestrial planet formation. Subsequent generations of small bodies may comprise S-type asteroids, whose formation region is proposed to be around 2 a.u. based on their isotopic affinity with Mars. S-type asteroids have a bimodal distribution (LL-like and H-like) and their accretion time is about 2-4 Ma according to associated meteorites (OCs). They do not show strong evidence of aqueous alteration and thus likely formed inward of the snowline. Ch- and Cgh-types have been placed further away from the Sun, namely beyond the snowline, since their associated meteorites (CM) show extensive evidence of aqueous alteration. CMs also contain chondrules (10-35% in volume) and their parent bodies thus likely formed relatively 'close' to those of S-type asteroids. The B,C, Cb and Cg types have been positioned in the Jupiter / Saturn region. They are expected to have accreted further than Ch- and Cgh-types, since they do not show any evidence of heating, and their fine-grained nature implied by their connection with IDPs (<1 µm from their emissivity features) testifies of the absence of a significant amount of chondrules in these bodies. Their density is also lower than that of Ch- and Cgh-types, implying a higher content of volatiles/ices. Still, their composistion (similar to pyroxene-rich IDPs) is different from that of P- and D-types or comets (olivine-rich IDPs). This compositional gradient (pyroxene-rich material lying closer to the star whereas the olivine-rich material is located further away from the star) is coherent with observations and models of accretion disks. The exact position of the limit between pyroxene-rich and olivine-rich remains to be constrained. The limit between grey-KBO and red-KBO has been placed at 20 a.u. following the predicted position of the methanol snowline.

**(c)** From the construction of (b) it is possible to draw the migration path towards (a) of the different compositional classes of small bodies. For the inner solar system, this exercise reveals that only minor outward transfers of object are required, while a major inward transfer is necessary from the outer solar system reservoirs. Outward transfers are needed in the outer solar system, in order to generate the excited population of KBOs.

**(d)** Popular migration models of giant planets (Nice and Grand Tack models) and the implied migration direction of the different compositional classes of small bodies resulting from the proposed giant planet' travel.